\documentclass[10pt, letterpaper]{article}
\usepackage{amsfonts,amsmath,amssymb,mathrsfs,verbatim}

\let\n\noindent

\newcommand{\btab}{\begin{tabular}}     \newcommand{\etab}{\end{tabular}}

\newcommand{\bt}{\begin{table}}     \newcommand{\et}{\end{table}}

\newcommand{\ba}{\begin{array}}     \newcommand{\ea}{\end{array}}
\newcommand{\bc}{\begin{center}}        \newcommand{\ec}{\end{center}}
\newcommand{\bfig}{\begin{figure}}      \newcommand{\efig}{\end{figure}}
\newcommand{\bp}{\begin{picture}}       \newcommand{\ep}{\end{picture}}
\newcommand{\bq}{\begin{quote}}     \newcommand{\eq}{\end{quote}}
\newcommand{\ben}{\begin{enumerate}}    \newcommand{\een}{\end{enumerate}}

\font\tenmsy=msbm10
\font\sevenmsy=msbm10 at 7pt
\font\fivemsy=msbm10 at 5pt
\newfam\msyfam 
\textfont\msyfam=\tenmsy
\scriptfont\msyfam=\sevenmsy
\scriptscriptfont\msyfam=\fivemsy
\def\blackB{\fam\msyfam\tenmsy}
\def\Z{{\blackB Z}}

\let\d\partial

\let\s\sigma

\let\R\rangle

\def\frac#1#2{{\textstyle{#1\over #2}}}

\def\text#1{\quad\hbox{#1}\quad}

\def\la{\lambda}

\def\ka{\kappa}
\def\de{\delta}

\def\rw{\rightarrow}

\def\R{\rangle}

\def\frac#1#2{{#1 \over #2}}

\let\d=\partial

\def\M{{\cal {M}}}

\def\rw{{\rightarrow}}

\let\d\partial

\let\s\sigma

\let\R\rangle

\def\frac#1#2{{\textstyle{#1\over #2}}}

\def\text#1{\quad\hbox{#1}\quad}

\def\la{\lambda}

\def\rw{\rightarrow}

\def\R{\rangle}

\def\frac#1#2{{#1 \over #2}}

\let\d=\partial
\def\S{{\cal S}}

\def\rw{{\rightarrow}}

\begin{document}

\vskip18pt

\title{\vskip60pt Embedding of bases: from the $\M(2,2\ka+1)$  to the $\M(3,4\ka+2-\delta)$ models}


\vskip18pt

\smallskip
\author{ P. Jacob and P.
Mathieu\thanks{patrick.jacob@durham.ac.uk,
pmathieu@phy.ulaval.ca.} \\
\\
Department of Mathematical Sciences, University of Durham, Durham, DH1 3LE, UK\\
and\\
D\'epartement de physique, de g\'enie physique et d'optique,\\
Universit\'e Laval,
Qu\'ebec, Canada, G1K 7P4.
}

\vskip .2in
\bigskip
\date{August 2005}

\maketitle


\vskip0.3cm
\centerline{{\bf ABSTRACT}}
\vskip18pt

A new quasi-particle basis of states is presented  for all the irreducible modules of the $\M(3,p)$ models. It is formulated in terms of a combination of Virasoro modes and the modes of the field $\phi_{2,1}$. This leads to a fermionic expression for particular combinations of  irreducible $\M(3,p)$ characters, which turns out to be identical with the previously known formula. Quite remarkably, this new quasi-particle basis embodies a sort of embedding, at the level of bases, of the minimal models  $\M(2,2\ka+1)$  into the $\M(3,4\ka+2-\delta)$ ones, with $0\leq \delta\leq 3$.


\n

\vskip6.5cm

\hbox{DCPT-05/XX}

\newpage



The $\M(3,p)$ models have been reformulated recently  \cite{JM3p} in terms of the extended
algebra defined by the OPEs
\begin{equation}\label{exal}
\begin{split}
\phi(z)\phi(w)=& {1\over (z-w)^{2h}} \left[ I+(z-w)^2 {2h\over c} T(w)+\cdots \right]\S \;, \cr
T(z) \phi(w) = & {h\phi(w)\over (z-w)^2}+{\d\phi(w)\over (z-w)}+\cdots \cr
T(z) T(w) = & {c_{3,p}/2\over (z-w)^4}+ {2T(w)\over (z-w)^2}+{\d T(w)\over (z-w)}+\cdots
\end{split}
\end{equation}
with
\begin{equation}
\phi\equiv \phi_{2,1}\;, \qquad h\equiv h_{2,1}= {p-2\over 4}\; ,
\qquad c_{3,p}= 1-2\frac{(3-p)^2}{p}\;, \end{equation} and
$\S=(-1)^{p{\cal F}}$ where ${\cal F}$ counts the number of $\phi$
modes. The highest-weight states $ |\s_\ell\R$ are completely
characterized by an integer $\ell$ such that $0\leq \ell\leq
(p-2)/ 2$ and satisfy
\begin{equation}\label{hwc}
\phi_{-h-n+\frac\ell2}\, |\s_\ell\R= 0 \qquad n>0\;.
\end{equation}
The highest-weight  modules  are described by the successive action of the lowering $\phi$-modes subject to exclusion-type constraints.
In the $N$-particle sector, with strings of lowering modes written in the form \cite{JM3p}\footnote{In \cite{JM3p}, the conditions are formulated in terms of the indices $n_i$ defined by:
$$
\phi_{-h+\frac\ell2+\frac{(N-1)}2-n_1} \,
\phi_{-h+\frac\ell2+\frac{(N-2)}2-n_{2}} \cdots
 \phi_{-h+\frac\ell2+\frac12-n_{N-1}}\,
  \phi_{-h+\frac\ell2-n_N}\, |\s_\ell\R\; .
$$ The relation between $s_i$ and $n_i$ is
$$s_i= n_i+h-\frac\ell2-\frac{(N-i)}2\;.$$}  (see also \cite{FJM,FJMMT} for $\ell=0)$:
\begin{equation} \label{strut}
  \phi_{-s_1} \,
\phi_{-s_{2}} \cdots
 \phi_{-s_{N-1}}\,
  \phi_{-s_N}\, |\s_\ell\R\;,
\end{equation}
these constraints are
\begin{equation}\label{jagG}
 s_i\geq s_{i+1}-2h+2\;, \qquad
   s_i\geq s_{i+2}+1\;, \qquad  s_{N-1} \geq -h+\frac\ell2+1\;,  \qquad   s_N\geq h-\frac\ell2\;, \end{equation}
 with
   \begin{equation}\label{zco}
  s_{N-2i} \in\Z+h+\frac\ell2\qquad {\rm and }\qquad   s_{N-2i-1}\in\Z-h+\frac\ell2\;.\end{equation}
   The complete module of $|\s_\ell\R$ is obtained by summing over all these states (\ref{strut}) satisfying (\ref{jagG}) and  all values of $N$.  The enumeration of these states lead to the standard form of the fermionic character for the sum of the two Virasoro modules $|\phi_{1,\ell+1}\R$ and $ |\phi_{1,p-\ell-1}\R$ of the $\M(3,p)$ models \cite{Byt, Wel, FFW} when $0\leq \ell\leq [p/3]$ (the closed form expression of the generating functions has not been obtained for the remaining cases).

Here we display a new form of the basis of states of the $\M(3,p)$ models, still viewed form the point of view of the extended algebra (\ref{exal}).
This basis is written in terms of combined sequences of Virasoro and $\phi$ modes, as
\begin{equation}\label{seq}
 L_{-n_1} \cdots L_{-n_N}{\phi}_{-m_1}  \cdots{\phi}_{-m_{M}} |\sigma_\ell \rangle \;.  \end{equation}
The module over $|\sigma_\ell \rangle $ is again the direct sum of the two Virasoro modules
 $|\phi_{1,\ell+1}\R$ and $ |\phi_{1,p-\ell-1}\R$.   In order to specify the constraints on the mode indices, we first define two integers $\ka$ and $\delta$ through the decomposition of $p$ as
\begin{equation}
p=4\ka+2-\delta\;, \qquad {\rm where}\qquad 0\leq \delta\leq 3\;.
\end{equation}
The conditions take the form
\begin{equation}\label{didi}
 n_i \geq n_{i+1}  \;, \qquad n_i \geq n_{i+\ka-1} + 2 \;, \qquad
m_i \geq m_{i+1} +\frac\delta2\;, \qquad m_i \geq m_{i+2} +\ka\;.
\end{equation}
These are supplemented by  the boundary conditions:
\begin{equation}\label{bdry}
 n_{N-\ell}\geq 2\;,  \qquad n_N\geq M+2-{\rm min}(\ell,1)\;,  \qquad  m_{M-1}\geq h-\frac\ell2 +{\rm max}(0,\ell-\ka)+\frac{\delta}{2} \;,  \qquad   m_M\geq h-\frac\ell2 \;.
\end{equation}
The $n_i$ are always integers but the range of the  indices $m_i$ is defined as follows. Given that $h= -\delta/4$ mod 1,  we have
 \begin{equation}
 m_{N-2i}\in \Z-\frac{\delta}4-\frac\ell2\qquad {\rm and }\qquad m_{N-2i-1}\in \Z+\frac\delta4-\frac\ell2\;,
 \end{equation} which are actually equivalent  to the conditions on the $s_i$ in (\ref{zco}).

The conditions (\ref{didi}) indicates that the Virasoro modes are ordered and further subject to a difference 2 condition at distance $\ka-1$. The $\phi$ modes are also ordered, being in fact all distinct if $\delta\not=0$. In addition, they are subject to a difference $\ka$ condition at distance $2$ (which is almost the `dual' of the conditions on the $n_i$).

The different inequalities in the boundary conditions (\ref{bdry}) have the following interpretation. At first, $m_M\geq h-\frac\ell2$ is simply the highest-weight condition (\ref{hwc}). The condition on $m_{M-1}$  partially specifies the different descendant states according to the value of $\ell$.  It is analogous to the third condition in (\ref{jagG}).  The inequality $n_{N-\ell}\geq 2$ means that the maximal number of $L_{-1}$ modes that can appear in the descendants of the $|\sigma_\ell\R$ module is $\ell$. Actually, this number is also bounded by the difference 2 condition at distance $\ka-1$, so that this maximal number is actually min $(\ell,\ka-1)$.

The most interesting condition is the remaining one in (\ref{bdry}), which, for the vacuum module $(\ell=0)$ reads $n_N\geq M+2$. For $M=0$, this takes into account the Virasoro highest-weight condition on the vacuum. But if there are $M$ $\phi$-modes already acting on the highest-weight state, the condition implies that all the modes $L_{-n}$ with $2\leq n\leq M+1$ have to be  excluded.  This can be interpreted as a sort of repulsion between the $T$ and $\phi$ `quasi-particles'. For any other module ($\ell\not=0)$, the bound  on $n_N$ reads $n_N\geq M+1$.

If $\ka=1$, the difference condition on the Virasoro modes becomes $n_i\geq n_i+2$, which is impossible. This means that when $\ka=1$, there can be no  Virasoro modes; the basis is solely described by the $\phi$ modes. Let us check that it reduces then to basis (\ref{jagG}). When $\ka=1$, $p=6-\delta$, but  in order for $p$ to be relatively prime with 3, we require $\delta=1$ or $2$. For $\delta=2$, so that $p=4$, the conditions (\ref{didi}) reduce to
$m_i\geq m_{i+1}+1$, in agreement with (\ref{jagG}) (note that the condition $m_i\geq m_{i+2}+1$ is thus automatically satisfied).  In that case $h=1/2$ and this
indeed describes the free-fermionic basis of the Ising model.
For $p=5$, these conditions take the form
$m_i\geq m_{i+1}+1/2$, which again implies the condition at distance 2. This agrees with (\ref{jagG}) and the known quasi-particle basis formulated in terms of the graded parafermion of dimension $h=3/4$
(cf. \cite{JM}, end of section 5, and \cite{JM3p} section 1.4).

To illustrate further these conditions, we present two examples in more detail. First we consider the $\M(3,8)$ model, so that $\ka=\delta=2$, and $h=3/2$.  Let us focus on the Virasoro vacuum module which corresponds to $\ell=0$ and which involves only  those descendant states that contain an even number of $\phi$ modes. The main (bulk)  conditions are
\begin{equation}\label{didi8}
 n_i \geq n_{i+1} + 2 \;, \qquad
m_i \geq m_{i+1} +1\;, \qquad m_i \geq m_{i+2} +2\;,
\end{equation}
(the last condition being in fact irrelevant here), while the  boundary conditions are simply $n_N\geq 2+M$ and $m_M\geq 3/2$. Let us denote the states in (\ref{seq}) by the combination of the two partitions $(n_1,\cdots, n_M;\, m_1\cdots , m_M)$. At the first few $(\leq 8)$ levels, the states are
\begin{equation}
\begin{split}
& 2:\quad (2;)\cr
& 3:\quad (3;)\cr
& 4:\quad (4;)\, (;\tfrac52,\tfrac32)\cr
& 5:\quad  (5;)\, (;\tfrac72,\tfrac32)\cr
& 6:\quad  (6;)\, (4,2;)\, (;\tfrac92,\tfrac32)\, (;\tfrac72,\tfrac52)\cr
& 7:\quad  (7;)\, (5,2;)\, (;\tfrac{11}2,\tfrac32)\, (;\tfrac92,\tfrac72)\cr
& 8:\quad  (8;)\, (6,2;)\, (5,3;)\, (;\tfrac{13}2,\tfrac32)\, (;\tfrac{11}2,\tfrac52)\, (;\tfrac92,\tfrac52)\, (4;\tfrac52,\tfrac32)
\end{split}
\end{equation}
The state $(4;\tfrac52,\tfrac32)$, which describes  $L_{-4}\phi_{-5/2}\phi_{-3/2}|\sigma_0\R$, is the first state involving both type of modes. Within this module, the first state with four $ \phi$ factors arises at level 12 and it is $(;\tfrac92, \tfrac72, \tfrac52, \tfrac32)$.
Similarly the first term with two Virasoro modes and two $\phi$ modes is $(6,4;\, \tfrac52, \tfrac32)$, while
the first one with three  Virasoro modes is   $(6,4,2;)$. This counting of states is to be compared with the expansion of the Virasoro character $\chi_{1,1}^{(3,8)}(q)$ (all the characters being normalized such that $\chi(0)=1$). Note that the $\M(3,8)$ model is equivalent to the superconformal minimal model ${\cal SM}(2,8)$.  Within the latter context, the above basis mixes the $G=\phi$ and $L$ modes.

For our second example, consider the $\M(3,14)$ model and the module with $\ell=4$. Here $\ka=3, \, \delta=0$ and $h=3$,  so that (\ref{didi}) takes the form
\begin{equation}\label{didib}
 n_i \geq n_{i+2} + 2 \;, \qquad
m_i \geq m_{i+1}\;, \qquad m_i \geq m_{i+2} +3\;,
\end{equation}
with the  boundary conditions
\begin{equation}
n_N\geq 1+M\;, \qquad n_{N-4}\geq 2\;, \qquad m_M\geq 1\;, \qquad m_{M-1}\geq 2\;.
\end{equation}
 At the first few $(\leq 6)$ levels, those states that contain an even number of $\phi$ modes, which pertains to the Virasoro module $|\phi_{1,5}\R$,  are
\begin{equation}
\begin{split}
& 1:\quad (1;)\cr
& 2:\quad (2;)\, (1,1;)\cr
& 3:\quad (3;)\, (2,1;)\, (;2,1)\cr
& 4:\quad (4;)\, (3,1;)\, (2,2;)\, (;3,1)\, (;2,2)\cr
& 5:\quad  (5;)\, (4,1;)\, (3,2;)\, (3,1,1;)\, (;4,1)\, (;3,2)\cr
& 6:\quad  (6;)\, (5,1;)\, (4,2;)\, (3,3;)\, (4,1,1;)\,(3,2,1;)\,  (;5,1)\, (;4,2)\, (;3,3)\, (3;2,1)\cr
\end{split}
\end{equation}
There are no terms containing  $L_{-1}^3$  because min $(\ell,\ka-1)=2$. Similarly, the state $\phi_{-1}\phi_{-1}|\sigma_4\R$ is excluded by the boundary condition on $m_{M-1}$. The first state with four $\phi$ modes is $(;5,4,2,1)$, at level 12 and the first state with two copies of both types of modes is $(3,3;2,1)$ at level 9.  The counting of states agrees with that coded in the Virasoro character $\chi_{1,5}^{(3,14)}(q)$. If we also allow states with an odd number of $\phi$ modes, we get instead the sum of  Virasoro characters $\chi_{1,5}^{(3,14)}(q)+ q\,  \chi_{1,9}^{(3,14)}(q)$. Note that $\M(3.14)\simeq W_3(3,7)$, so that the above is an example of a $W_3$ basis involving both the $T$ and $W$ modes.

Let us stress a remarkable feature of the new basis. The conditions (\ref{didi}) for the Virasoro modes  are precisely the one pertaining to the quasi-particle basis of the $\M(2,p)$ models, with $p=2\ka+1$ \cite{FNO}. Moreover, the boundary condition on $n_{N-\ell}$, which specifies the maximal number of $L_{-1}$ factors, thereby distinguishing the different modules, is also the very one that occurs in these models. Therefore, in absence of $\phi$ modes, the above $\M(3,4\ka+2-\delta)$ basis reduces  to the  $\M(2,2\ka+1)$ one. It thus appears that the above basis describes a sort of embedding of the $\M(2,2\ka+1)$ models within  the $\M(3,4\ka+2-\delta)$ ones.


Let us consider  the expression for the  characters associated to this new basis. Constructing these characters amounts to finding  the generating function for  the composition of the two partitions $(n_1,\cdots , n_N)$ and $(m_1,\cdots, m_M)$ satisfying (\ref{didi}) and (\ref{bdry}).
This is essentially built from the composition of two corresponding generating functions, both of which being known (up to a restriction on $\ell$ to be specified).

The generating functions for partitions  $(n_1,\cdots , n_N)$ is obtained as follows. First, delete $M$ from each parts $n_i$ and introduce $q^{NM}$ to correct for this. The resulting restricted partitions are enumerated  by the Andrews multiple-sum \cite{And, Andy}:
\begin{equation}
{\cal H}_{\kappa,\ell}(q)z^N
= \sum_{s_i\geq 0, \atop \sum  i s_i= N}{ q^{N_1^2+\cdots+
 N_{\ka-1}^2+N_{\ell+1}+\cdots N_{\ka-1}+NM} z^N \over (q)_{s_1}\cdots (q)_{s_{\ka-1}} }\;,
\end{equation}
where
 $N_i= s_i+\cdots +s_{\ka-1}$.

Similarly, the generating function
for partitions  $(m_1,\cdots , m_M)$ can be extracted from  \cite{FJM} up to simple modifications. The latter   generating function enumerates the partitions $(\la_1,\cdots , \la_M)$ satisfying
\begin{equation}
\la_i\geq \la_{i+1}\;, \qquad \la_i\geq \la_{i+2}+2r\;, \qquad  \la_M\geq 1\;,  \qquad  \la_{M-1}\geq 1+  {\rm max}\; (0,{\tilde \ell}-1)\;,
\end{equation}
for $0\leq {\tilde \ell} \leq [(2r+5)/3]$ where $[x]$ stands for the integer part of $x$ (the boundary condition of $\la_{M-1}$ induces a correcting term  in the generating function
that has been introduced in \cite{JM3p}.)
To connect the two problems, let us redefine $m_i$ as:
\begin{equation}
m_i= \la_i+h-\frac\ell2-1+({M-i})\frac\delta2\;.
\end{equation}
The conditions (\ref{didi})-(\ref{bdry}) become then
\begin{equation}
\la_i\geq \la_{i+1}\;, \qquad \la_i\geq \la_{i+2}+\ka-\delta\;, \qquad\la_M\geq1\;, \qquad \la_{M-1}\geq 1+ {\rm max}\; (0, \ell-\kappa)\;.
\end{equation}
We thus recover the counting problem of \cite{FJM,JM3p} but with $2r\, \rw\, \ka-\delta$ and ${\tilde \ell}-1 \, \rw \,\kappa-\ell$.  (Note that the generating function of  \cite{FJM} does not hold for those cases where $2r+5$ is divisible by 3. But this is not restrictive since if $\ka-\delta+5$ were a multiple of 3, say $3n$, then $p$ would be $12n+3\delta-18$, which is divisible by 3 and that would not corresponds to a minimal model.)
The correcting factor
$q^{M(M-1)\frac\de4+M(h-\frac\ell2-1)}$ will keep track of the shifted staircase that must be added to adjust the weight when passing from  the  partitions $(\la_1\cdots, \la_M)$ to our original partitions $(m_1,\cdots, m_M)$. From \cite{FJM, JM3p}, we see that the generating function is written as a $g$-multiple sum, where $g$ is given by
\begin{equation}
g= \left[{\kappa-\delta +5\over 3}\right]\;.
\end{equation}
and it takes the form
\begin{equation}{\cal G}_{g,\ell}(q)z^M= \sum_{t_1,t_2,\cdots ,t_g\geq 0\atop 2(t_1+\cdots+ t_{g-1}) + t_g= M} {q^{tBt + Ct+M(M-1)\frac\de4+M(h-\frac\ell2-1)}
\; z^{M}\over (q)_{t_1}\cdots (q)_{t_g}}\
\end{equation}
(with the understanding that
$ tBt= \sum_{i,j=1}^g t_i \, B
_{ij}\,  t_j$ and $ Ct=
\sum_{i=1}^g {C
}_i\,t_i$), and
the $g\times g$  symmetric  matrix $B$ reads
\begin{equation}
B
= \begin{pmatrix}\ka-\de&\ka-\de&\cdots &\ka-\de &\frac{\ka-\de}2\cr \ka-\de&\ka-\de+1&\cdots &\ka-\de+1 &\frac{\ka-\de}2+\frac12\cr
\cdots&\cdots
&\cdots&\cdots &\cdots\cr
\ka-\de&\ka-\de+1 &\cdots &\ka-\de+g-2&\frac{\ka-\de}2-1+\frac{g}2\cr
 \frac{\ka-\de}2& \frac{\ka-\de}2+\frac12 &\cdots
&\frac{\ka-\de}2-1+\frac{g}2& g-1
\cr
\end{pmatrix}\;,
\end{equation}
while the entries of the row matrix $C$ are
\begin{equation}C
_j=-\ka+\delta +j+1+{\rm max}\, (0,\ell-\ka)\quad {\rm for}\quad j<g \qquad {\rm and} \qquad C
_g=-g+2\;.
\end{equation}
We stress that this result holds only for $0\leq \ell\leq \ka+g-1$ (and this range is identical to the previous one $0\leq \ell\leq [p/3]$ since $3\ka+3g-3=p$).  For the remaining values of $\ell$, that is, for $\ka+g \leq \ell \leq p/2-1$, we stress that although the generating function  has not been found in closed form, the validity of the basis has been verified to high order in $q$.

The composition of these two generating functions is obtained by the multiplication of ${\cal H}_{\kappa,\ell}(q)z^N $ with ${\cal G}_{g,\ell}(q)z^M$, setting $z=1$, and summing over $N$ and $M$. This leads to the expression
\begin{equation}
\chi_{\ell}^{(3,4\ka+2-\de)}(q)= \sum_{s_1,\cdots , s_{\ka-1}, t_1,\cdots,  t_g\geq 0}{ q^{N_1^2+\cdots+
 N_{\ka-1}^2+NM+N_{\ell+1}+\cdots N_{\ka-1}+tBt + Ct+M(M-1)\frac\de4+M(h-\frac\ell2-1)} \over (q)_{s_1}\cdots (q)_{s_{\ka-1}} (q)_{t_1}\cdots (q)_{t_{g}} }
\end{equation}
Now, by redefining the summation variables as
\begin{equation}
(s_1,\cdots, s_{\ka-1},t_1,\cdots, t_g)=(n_1,\cdots, n_{\ka+g-1}) \;,
\end{equation}
 we can reexpress the above character in the compact  form:
\begin{equation}
\chi_{\ell}^{(3,4\ka+2-\delta)}(q)=
\sum_{n_1,\cdots,  n_{\ka+g-1} \geq 0}  { q^{n{\cal B} n + {\cal C}n }
\over (q)_{n_1}\cdots (q)_{n_{\ka+g-1}}} \;,
\end{equation}
where the matrices ${\cal B}$ and ${\cal C}$ are  defined as follows, with $1\leq i,j\leq \ka+g-2$:
\begin{equation}
 {\cal B}_{i,j}= {\rm min} (i,j) \; , \qquad
  {\cal B}_{j,\ka+g-1}= {\cal B}_{\ka+g-1,j} = \frac{j}2\; , \qquad  {\cal B}_{\ka+g-1,\ka+g-1}= g-1+{\delta\over 4}\;,
\end{equation}
and
\begin{equation}
{\cal C}_j= {\rm  max}\; (j-\ell,0)\;, \qquad {\cal C}_{\ka+g-1}= \ka+1-g-{\delta\over 2}\;.
\end{equation}
This is the form obtained in \cite{Byt, Wel, FFW, JM3p}. This in turn demonstrates the correctness of the basis, at least for $\ell\leq \ka+g-1$.
As previously indicated, this character is equal to the  following sum of Virasoro characters:
\begin{equation}
\chi_{\ell}^{(3,p)}(q)= \chi_{1,\ell+1}^{(3,p)}(q)+ q^{h-\ell/2}\, \chi_{1,p-\ell-1}^{(3,p)}(q)\;.
\end{equation}

\noindent {\bf ACKNOWLEDGMENTS}

The work of PJ is supported by EPSRC and partially  by the EC
network EUCLID (contract number HPRN-CT-2002-00325), while that of  PM is supported  by NSERC.

\end{document}